\documentclass[conference]{IEEEtran}
\IEEEoverridecommandlockouts
\usepackage{cite}
\usepackage[bookmarks=false]{hyperref}
\usepackage{amsmath}
\usepackage{amssymb}
\usepackage{amsfonts}
\usepackage{algorithmic}
\usepackage{graphicx}
\usepackage{textcomp}
\usepackage{xcolor}
\usepackage{multirow}
\usepackage{fancyhdr}
\usepackage{booktabs}
\usepackage{colortbl}
\usepackage[labelfont=bf]{caption}
\usepackage{subfig}
\def\BibTeX{{\rm B\kern-.05em{\sc i\kern-.025em b}\kern-.08em
    T\kern-.1667em\lower.7ex\hbox{E}\kern-.125emX}}
 
\usepackage[left=1.5cm,top=1.35cm,right=1.5cm,bottom=1.6cm,nofoot]{geometry}

\begin{document}

\title{One-shot Learning for iEEG Seizure Detection Using End-to-end Binary Operations: Local Binary Patterns with Hyperdimensional Computing
\thanks{Support was received from the ETHZ Postdoctoral Fellowship Program, the Marie Curie Actions for People COFUND Program, and the EU Mnemosene.}
}

\author{\IEEEauthorblockN{Alessio Burrello\IEEEauthorrefmark{1}, Kaspar Schindler\IEEEauthorrefmark{2}, Luca Benini\IEEEauthorrefmark{1}, Abbas Rahimi\IEEEauthorrefmark{1}}

\IEEEauthorblockA{\IEEEauthorrefmark{1}Integrated Systems Laboratory, ETH Zurich, Switzerland 
\IEEEauthorrefmark{2}Sleep-Wake-Epilepsy-Center, Inselspital Bern, Switzerland}
\IEEEauthorblockA{Emails: bualessi@student.ethz.ch,  kaspar.schindler@insel.ch, lbenini@iis.ee.ethz.ch,
abbas@iis.ee.ethz.ch}
}
%\IEEEauthorrefmark{3}
%\IEEEauthorrefmark{2}
\maketitle

\begin{abstract}
This paper presents an efficient binarized algorithm for both learning and classification of human epileptic seizures from intracranial electroencephalography (iEEG).
The algorithm combines local binary patterns with brain-inspired hyperdimensional computing to enable end-to-end learning and inference with binary operations. 
The algorithm first transforms iEEG time series from each electrode into local binary pattern codes. Then \emph{atomic} high-dimensional binary vectors are used to construct composite representations of seizures across all electrodes. 
%
%Such binary operations are far more efficient than floating-point counterparts. 
%particularly important for implantable devices.
For the majority of our patients (10 out of 16), the algorithm quickly learns from one or two seizures (i.e., one-/few-shot learning) and perfectly generalizes on 27 further seizures.
For other patients, the algorithm requires three to six seizures for learning.
Overall, our algorithm surpasses the state-of-the-art methods~\cite{LGP+MLP} for detecting 65 novel seizures with higher specificity and sensitivity, and lower memory footprint.
\end{abstract}

%\begin{IEEEkeywords}
%iEEG, one-shot learning, local binary patterns, symbolic dynamics, hyperdimensional computing
%\end{IEEEkeywords}

\section{Introduction}
\thispagestyle{fancy}
\fancyhf{}
\chead{Published as a conference paper at the IEEE BioCAS 2018}

Epilepsy is a chronic neurological disorder affecting 0.6--0.8\% of the world's population~\cite{Epilesy_intro}. 
One third of patients with epilepsy continue to suffer from seizures despite pharmacological therapy~\cite{Seizure_diffusion}.
For these patients with drug-resistant epilepsy~\cite{Bern_Dataset}, efficient algorithms for seizure detection are needed in particular during pre-surgical long-term iEEG recordings.
iEEG currently provides the best spatial resolution and highest signal-to-noise ratio to record electrical brain activity. 

Most seizure detection studies assume that there are two
distinct states of brain activity in patients with epilepsy (i.e., interictal and ictal), and that these states can be detected by iEEG.
One major challenge then is to reliably detect seizures from a small number of ictal examples.
The difficulties are due to the patient-specific nature of seizure dynamics, and to the asymmetry inherent in long-term iEEG, namely that the ratio of interictal to ictal segments is typically very large~\cite{paitSpecModel}.
This requires a fast algorithm that learns from few ictal iEEG segments and generalizes well for novel seizure recordings.

Another challenge is to realize such algorithm with low complexity suitable for execution on implantable devices for long-term operation.
One promising option is computing with simple linear binary codes to avoid otherwise expensive operations such as costly floating-point arithmetic.
Combining methods from symbolic dynamics and information
theory is a computationally efficient approach. At its core it consists of analyzing the occurrence of patterns and even bears similarity to classical visual EEG interpretation~\cite{Kaspar_redundancy}.
Local binary patterns (LBP)~\cite{LPB_1Dfeature}, as an elegant \emph{symbolization}, map a sequence of iEEG samples into a small bit string, depending solely on whether their amplitudes increase or decrease. 
These basic symbols can be further combined over time and across electrodes to generate a compact representation for encoding the state of interest.  
Such representations can be effectively constructed by using brain-inspired hyperdimensional (HD) computing~\cite{HD09} that offers the ability to learn object categories from one or few examples---also referred to as one-shot or few-shot learning---with simple distributed operations on long binary vectors~\cite{Rahimi2016,TCAS17,PULP_HD,BICT17}.

In this paper, we propose a new algorithm that jointly exploits LBP and HD computing to address the aforementioned challenges by the following contributions. %
First, the algorithm operates with end-to-end simple binary operations: (1) The LBP feature extractor directly transforms the time series into a short bit string, as a symbol, for every iEEG electrode. 
(2) HD computing then projects the bit strings to a high-dimensional vector to compute a holistic binary representation that encodes occurrences of the symbols among all electrodes. 
(3) The training and classification are performed by simply bundling and comparing the binary high-dimensional vectors.
(4) The classification decision is followed by a patient-dependent voting to reduce false alarms. 
Second, the algorithm quickly learns from one seizure, i.e. one-shot learning (for eight patients out of 16), or two seizures (for two more patients), and perfectly generalizes on detecting 27 novel seizures with $k$-fold cross-validation.
%
%The average delay of detection is 18.2\,s. enabling to raise an alarm in the first 17.9\% of mean seizure duration.
%
For the remaining six patients, the algorithm requires 3--6 seizures for learning.
Our algorithm surpasses state-of-the-art methods using local pattern transformation coupled with a linear support vector machine (SVM), and a multilayer perceptron (MLP) neural network~\cite{LGP+MLP}.
Furthermore, the algorithm is truly scalable and provides a simple interface (with minimal number of parameters) to universally operate with all patients having 36 to 100 electrodes implanted.
We provide the public access\footnote{Available for download at \url{http://ieeg-swez.ethz.ch}} to our anonymized dataset and codes.
\section{Background}
\subsection{Local Binary Pattern (LBP)}
\label{sec:LBP}
A class of data-analysis methods is referred to as symbolization, which describes the process of transforming raw experimental measurements into a series of discrete symbols. 
Symbolization is particularly interesting for EEG
analysis, because as recent experience has clearly demonstrated, it faithfully preserves dominant dynamical signal characteristics while significantly increasing the efficiency of detecting and quantifying information contained in real-world time series~\cite{symbolization}.
Symbolization may be efficiently achieved by mapping a sequence of iEEG samples into a bit string, i.e. a one-dimensional local binary pattern (LBP)~\cite{LPB_1Dfeature}.
A LBP code reflects relational aspects between consecutive values of the original iEEG signals only, but not the values themselves.

\begin{figure}
  \centering
\subfloat{\includegraphics[width=0.45\columnwidth]{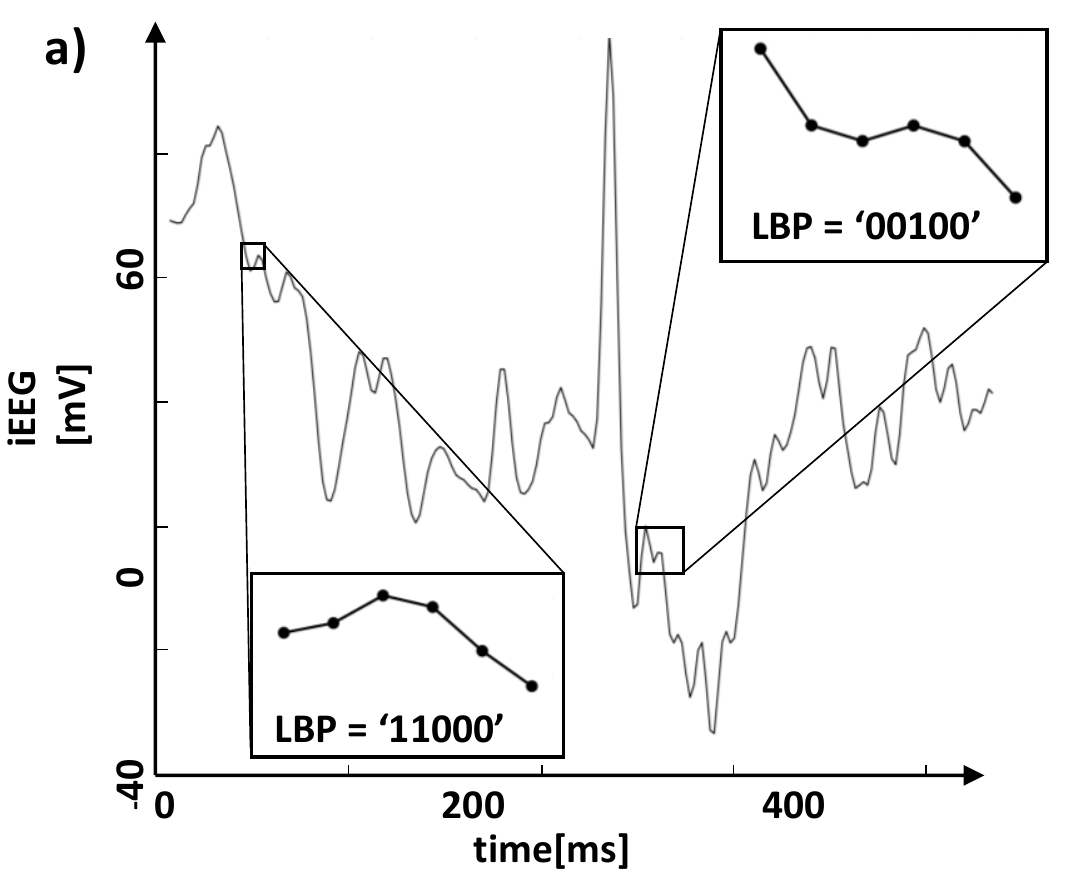}\label{fig:LBPint}}
  \hfill
\subfloat{\includegraphics[width=0.45\columnwidth]{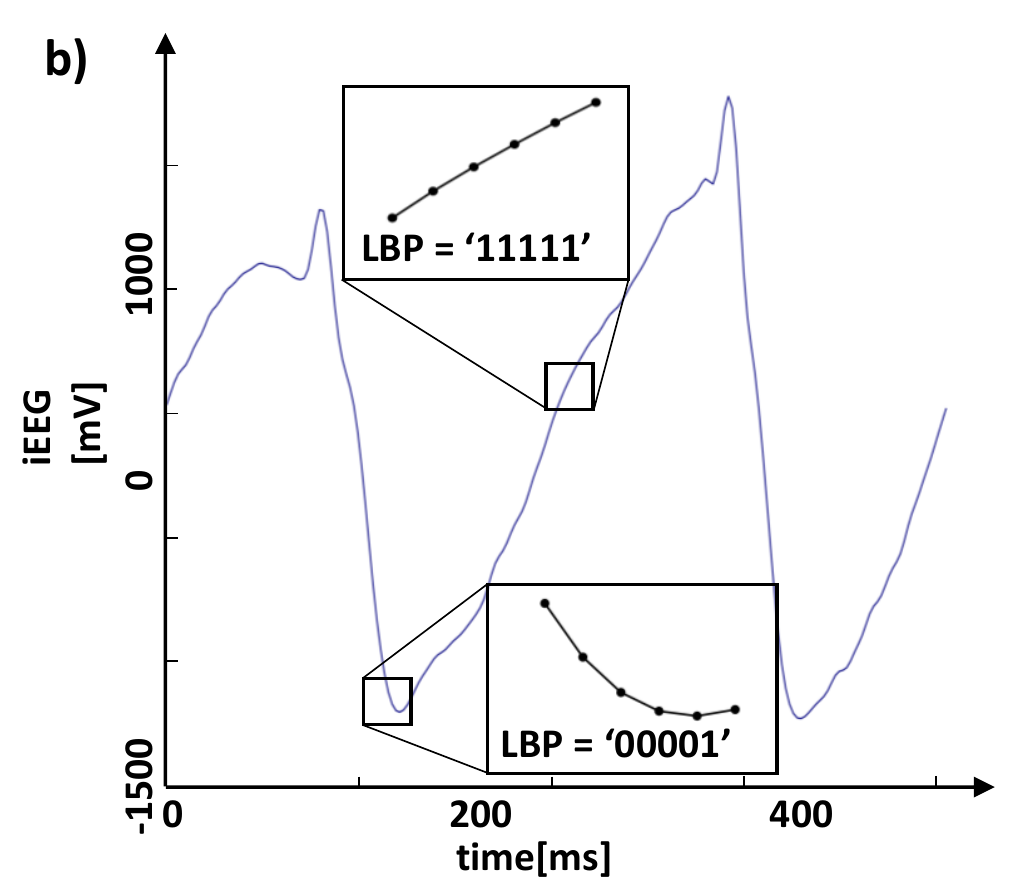}\label{fig:LBPict}}
  \hfill
\subfloat{\includegraphics[width=0.9\columnwidth]{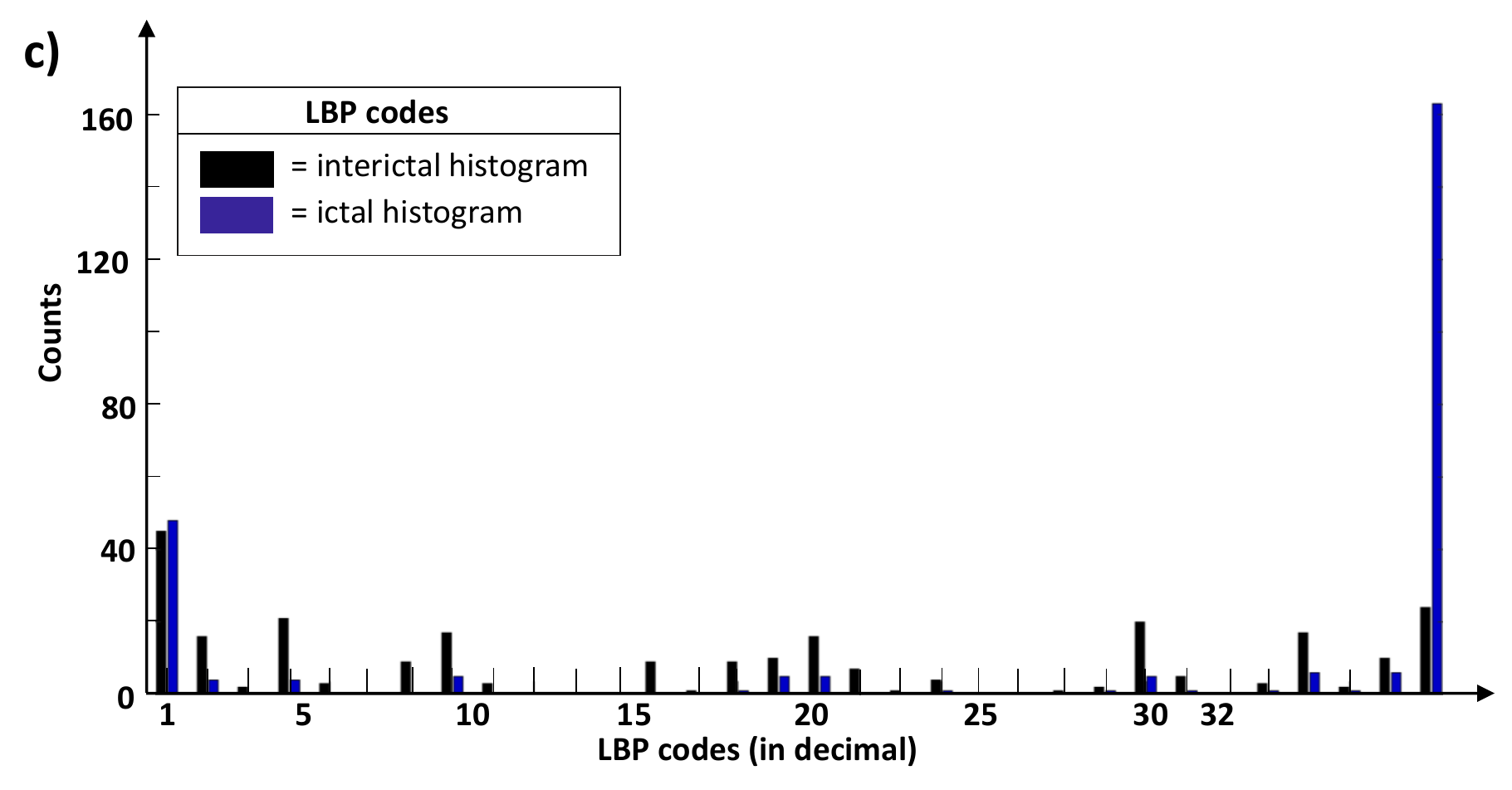}\label{fig:HISTint}}
  \hfill
  \hfill 
  \caption{iEEG signal during: \textbf{a)} interictal state, and \textbf{b)} ictal state, with examples of their LBP codes of $l$=5.
  \textbf{c)} Their corresponding histograms of LBP codes inside a 0.5\,s window.}
  
%  \caption{\textbf{a)} Local binary pattern (LBP) codes for $l=3$: a single point has a LBP code constructed by looking to the next 3 points; if a point is higher than the preceding, a $1$ is assigned, otherwise a $0$. \textbf{b)} iEEG signal during interictal state: the signal is non-stationary and the codes are well distributed over almost all the possible codes. \textbf{c)} iEEG in the middle of a seizure: the strong time-irreversible signal has a predominant portion of a single LBP code.}
\end{figure}

Computing a LBP code is simple:
(1) The iEEG signal samples are converted into a bit string depending on the sign of the temporal difference of adjacent samples. 
If the difference is positive, we assign a $1$ to the sampling point, otherwise a $0$.
(2) A LBP code of length $l$ is associated with every sampling point by concatenating its bit with the successive $l-1$ bits. 
Fig.~\ref{fig:LBPint}-\ref{fig:LBPict} show examples of LBP code with $l$=5.
Fig.~\ref{fig:HISTint} illustrates how histograms of LBP codes differ between interictal and ictal states.
During the interictal state the LBP codes are almost evenly distributed over all the possible codes. 
In contrast the ictal window has a predominant portion of a single LBP code and many LBP codes are missing due to the typically slow and asymmetric oscillations evolving during seizures. 
\subsection{Hyperdimensional (HD) Computing}
The human brain consists of billions of neurons, glial cells and synapses, suggesting that large circuits are fundamental to its computational power. 
HD computing explores this idea by invoking vectors of very high dimensionality for computing, i.e., $d$=10,000 dimensions~\cite{HD09}. 
There exists a huge number of different, nearly orthogonal vectors with the dimensionality in the thousands~\cite{Kanerva98SDM}.
This lets us combine two such vectors into a new vector using well-defined vector space operations, while keeping the information of the two vectors with high probability. 
HD computing has further unique features including fast learning, robustness, and efficiency of realization~\cite{TCAS17}.

HD computing has been used for classifying various biosignals including EEG~\cite{BICT17} and electromyography (EMG)~\cite{PULP_HD}.
Its learning and classification are composed of three main steps: 1) mapping symbols to \emph{atomic} high-dimensional vectors; 2) combining atomic vectors with well-defined arithmetic operations in an encoder to produce composite structural vectors; 3) storing/updating (i.e., learning) these vectors inside an associative memory and finally comparing with query vectors (i.e., inference).   
HD computing begins with selecting a set of random high-dimensional vectors (with i.i.d. components) to represent basic objects.
They serve as atomic vectors and are used as building blocks to construct representations of more complex objects. 
To generate these atomic vectors, we use random $d$-dimensional vectors of equally probable $1$s and $0$s, i.e., dense binary elements of $\lbrace 0, 1\rbrace^{d}$.
These vectors are stored to a so-called item memory (IM), i.e. a symbol table or dictionary of vectors defined in the system.
In our seizure detection system, the names of electrodes and the LBP codes are the basic symbols.
The IM assigns a random orthogonal vector to every symbol.

Here, we focus on two main operations of HD computing for encoding with the atomic vectors: bundling and binding.
Bundling, or addition, of binary vectors $[A + B + \ldots]$ is defined as the componentwise majority with ties broken at random.
Binding is defined as the componentwise XOR ($\oplus$).
Both operations produce a $d$-bit vector with an important distinction: bundling produces a vector that is \emph{similar} to the input vectors, whereas binding produces a \emph{dissimilar} vector.
Hence, bundling is well suited for representing sets, and can combine field/value bindings to produce a larger structure (e.g., record or tuple).
Representations of such composite structures are constructed directly from representations of the atomic vectors by applying these operations without requiring any learning for the encoder.

The output vector of the encoder is then fed into an associative memory (AM) for training and inference.
During training the output vector of the encoder is stored in the AM as a \emph{learned} pattern.
During inference the output of the encoder is compared with the learned patterns.
Comparison is based on a distance metric over the vector space. 
The AM uses Hamming distance, defined as the number of different components of two binary vectors.
\section{Proposed Algorithm: LBP Features and Binary HD Learning and Classification}
\begin{figure*}
  \centering
\includegraphics[width=0.95\textwidth]{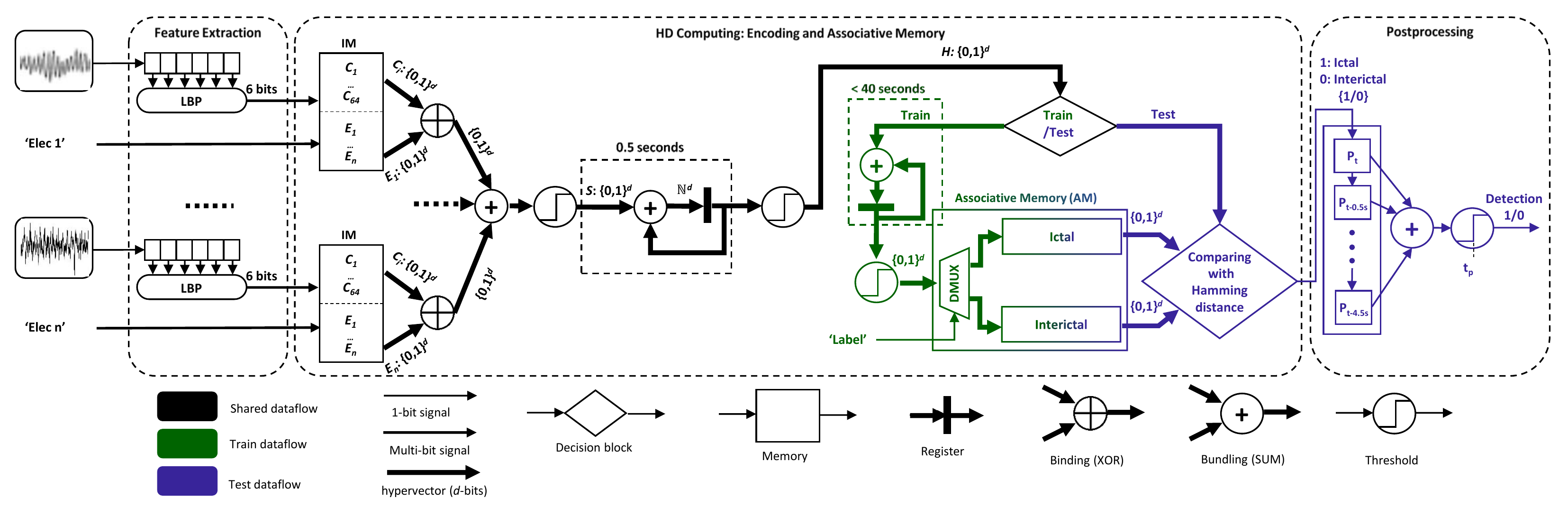}
  \caption{Binarized seizure processing chain: (1) Feature extraction generates a 6-bit LBP code for each electrode; (2) HD computing projects these codes into $d$-dimensional space and constructs vector $H$ that represents the histogram of 0.5\,s recording. During training the AM learns from this vector, and during inference provides a label for it; (3) As postprocessing, a simple patient-dependent ($t_p$) voting decides based on the last 10  labels.}
  \label{fig:architecture}
\end{figure*}
%
% \begin{figure}
%   \centering
% \includegraphics[width=1\columnwidth]{./Fig2b.PNG}
%   \caption{iEEG signal of a single channel. In green the window of points used to construct a single LBP symbol, shifted of 1 sample at a time (256 symbols inside a window of 0.5 seconds); in red the 0.5 seconds not-overlapping window used to construct the histogram; in blue the bigger time-window (5 seconds length, 0.5 seconds shifting) used to compute the final prevision.}
%   \label{fig:iEEG}
% \end{figure}
% 
The LBP feature extractor and HD computing can be combined to quickly learn from ictal iEEG to then detect further seizures. 
Our proposed algorithm uses LBP codes to directly symbolize the iEEG signal of an electrode.
Then a composite $d$-dimensional binary representation is constructed to capture the statistics of the LBP codes across all electrodes and over time.
The final classification is followed by simple postprocessing as shown in Fig.~\ref{fig:architecture}.
\subsection{Preprocessing and LBP Feature Extraction}
The iEEG signals are sampled by a 16-bit ADC, filtered by a fourth order Butterworth filter between 0.5\,Hz and 150\,Hz, and downsampled to 512\,Hz.
A LBP code with $l=6$ is computed for every sampling point. The LBP code considers six consecutive samples, and moves by one sample. 
Our LBP code generates $2^l$ different symbols that are fed into the next stage for learning and classification.
Using larger code sizes impairs its applicability to non-stationary signals and latency of classification.
%
%The code size determines the minimum size of following statistical analysis window: the size of such window should be large enough such that all symbols can at least theoretically occur once~\cite{Kaspar_redundancy}.   
%
%hence an analysis and classification window of 0.5\,s (containing 256 samples) is sufficient ($256>2^6$).
%
%
%
\subsection{HD Learning and Classification}
\label{sec:HD_Class}
HD computing first projects the LBP codes to the high-dimensional space via the IM that assigns an orthogonal vector to every LBP code, i.e., $C_1 \bot C_2 \dots \bot C_{64}$.
To combine these vectors across all electrodes, HD computing generates a spatial record ($S$), in which an electrode \emph{name} is treated as a field, and its LBP code is treated as the value of this field. 
The IM is also used to map the name of electrodes to orthogonal vectors: $E_1 \bot E_2 \dots \bot E_{n}$ for a patient with $n$ electrodes.
This allows to bind the name of each electrode ($E_j \mid j\in[1,n]$) to its corresponding code ($C_i \mid i\in[1,64]$).
The spatial record ($S$) is then constructed by bundling these bound vectors using majority gates: $S=[E_1 \oplus C_i + E_2 \oplus C_i + ... + E_n \oplus C_i]$.

Vector $S$ is computed for every new sample, and represents the spatial information about the LBP codes of all electrodes.
The next step is to compute the histogram of LBP codes for a moving window of 0.5\,s.
This window size should be large enough to theoretically permit at least a single occurrence of all possible LBP codes~\cite{Kaspar_redundancy,Forbidden_pattern}.
The window of 0.5\,s contains 256 LBP codes constructed with maximal overlap.
To have a high probability that every code occurs inside this window, we should hold $256>2^{l+1}$, hence $l<7$.
To estimate the histogram of LBP codes inside this window, a multiset of temporally generated $S$ vectors is computed as $H=[S^1+S^2+...+S^{256}]$.
A majority gate is applied in the temporal domain through accumulation (i.e., componentwise addition) of $S^t$ vectors $t \in \{1,...,256\}$, that are produced within the window, and then thresholding at half.  

We observe that the interictal and ictal states show different distributions of LBP codes inside the window: during an interictal segment, we have a nearly random signal, with a well distributed count histogram; conversely, during a seizure we typically observe rhythmic signals, i.e., slow and often temporally asymmetric oscillations, which yield polarized histograms as demonstrated in Fig.~\ref{fig:HISTint}.
This shows that the distribution of LBP codes, not necessarily their sequence, is an important indicator to distinguish ictal vs. interictal states. 
The high-dimensional space naturally encodes such histograms in $H$ by accumulating and thresholding the spatial vectors.
To reconstruct the histogram from $H$, we compare it with $C_1 \bot C_2 \dots \bot C_{64}$, inside the IM, and calculate the normalized Hamming distances that reflect relative frequency of the symbols.
The reconstructed histograms achieve a Pearson correlation coefficient $>0.9$ compared to the exact histograms.
The output of HD encoding is vector $H$, which is updated every 0.5\,s.
To quickly train the classifier, we use this vector to build the AM containing two \emph{prototype} vectors representing ictal and interictal labels.
To train the interictal prototype, all $H$ vectors computed over an interictal window of 40\,s are accumulated (summed), and then thresholded (binarized) to be stored in the AM. 
Correspondingly, an ictal prototype vector is generated from a smaller window of 10--30\,s depending on seizure duration. 
For classification, a newly computed $H$ vector is compared to every prototype of the AM using Hamming distance to determine its label.
\subsection{Postprocessing}
\label{sec:Postprocess}
The last part of the algorithm postprocesses the labels produced by the HD classifier every 0.5\,s.
It defines a window of 5\,s where a final decision is made based on the last 10 labels collected from the HD classifier (shifting labels of 0.5\,s at a time).
%
%This window size provides a trade off between the delay of detection and false alarms.
%
The decision is made based on a patient-specific threshold ($t_p$): the algorithm detects the seizure onset when the number of ictal labels inside the 5\,s window is equal or greater than $t_p$.
During the training, $t_p$ is initially set to 10 (out of 10, to reduce the false alarms) and is decreased such that the algorithm can detect the training ictal segment.
After the training, we obtain $t_p$ between 10 to 8 depending upon the patient.

Overall, our algorithm has five parameters: the size of LBP codes ($l$), the duration of the two windows (0.5\,s and 5\,s), $d$, and $t_p$.
Only the last parameter is patient-dependent, whereas the others are fixed for all patients.
Nevertheless, to reduce the memory load, $d$ can be adjusted to the individual patient depending on the number of electrodes and seizure dynamics. 
We observe that the algorithm works with $d$=10,000 for all patients. For some patients it may even be reduced to 1000 without impairing its performance.
\section{Dataset and Experimental Results}
\begin{table*}[]
\centering
\caption{Learning from one/two seizure(s) with perfect (100\%) generalization for patient majority using $k$-fold cross-validation.}
\label{table:Group1}
\begin{tabular}{ccclllcclcclcclcc}\toprule
\multicolumn{8}{c}{\textbf{Patients information}}                                                                                           & \multicolumn{3}{c}{\textbf{LBP + HD Computing}}                                 & \multicolumn{3}{c}{\textbf{LBP + Linear SVM}}                                & \multicolumn{3}{c}{\textbf{LGP + MLP}}                                       \\\midrule
\textbf{ID} & \textbf{\begin{tabular}[c]{@{}c@{}}Elect.\\ {[\#]} \end{tabular}} & \textbf{\begin{tabular}[c]{@{}c@{}}Seiz.\\ {[\#]}\end{tabular}} &\multicolumn{3}{c}{\textbf{\begin{tabular}[c]{@{}c@{}}Seizure duration {[}s{]}\end{tabular}}}& \textbf{\begin{tabular}[c]{@{}c@{}}Trained\\
seiz. {[\#]}\end{tabular}} & \textbf{\begin{tabular}[c]{@{}c@{}}$k$-fold \end{tabular}} & \textbf{\begin{tabular}[c]{@{}c@{}}Mean\\ delay {[}s{]}\end{tabular}} & \textbf{\begin{tabular}[c]{@{}c@{}}Spe.\\   {[}\%{]}\end{tabular}} & \textbf{\begin{tabular}[c]{@{}c@{}}Sen.\\   {[}\%{]}\end{tabular}} & \textbf{\begin{tabular}[c]{@{}c@{}}Mean\\ delay {[}s{]}\end{tabular}} & \textbf{\begin{tabular}[c]{@{}c@{}}Spe.\\   {[}\%{]}\end{tabular}} & \textbf{\begin{tabular}[c]{@{}c@{}}Sen.\\   {[}\%{]}\end{tabular}} & \textbf{\begin{tabular}[c]{@{}c@{}}Mean\\ delay {[}s{]}\end{tabular}} & \textbf{\begin{tabular}[c]{@{}c@{}}Spe.\\   {[}\%{]}\end{tabular}} & \textbf{\begin{tabular}[c]{@{}c@{}}Sen.\\   {[}\%{]}\end{tabular}} \\\cmidrule{4-6}
   &                      &                    & \textbf{Mean}                     & \textbf{Min} & \textbf{Max} &                          &                 &                    &                      &                      &                   &                     &                     &                   &                     &                     \\ \midrule
1& 100& 5& 14& 10& 22&  2& 4 &6.3&  100& 100&  6.9&100&100&    6.9&96.76&100\\
2& 64& 4& 146& 89& 179& 1& 4 &15.1& 100& 100&  10.1&91.74&75&  12.2&98.26&100\\
4& 42& 4& 223& 96& 301& 1& 4 & 34.5&100& 100&  29.3&100&100&   35.2&100&100\\
5& 59& 6& 88& 67& 117&  1& 6 &20.9& 100& 100&  14.7&92.09&100& 14.6&84.54&100\\
6& 36& 2& 15& 14& 16&   1& 2 &6.3&  100& 100&  9.0&100&100&    7.5&100&100\\
8& 61& 3& 121& 52& 184& 1& 3 &13.2& 100& 100&  11.9&100&100&   10.3&100&100\\
11& 59& 2& 57& 52& 61&  1& 2 &7.0&  100& 100&  6.5&100&100&    6.5&100&100\\
13& 98& 2& 99& 73& 125& 1& 2 &10.0& 100& 100&  16.3&100&100 &  9.8&100&100\\
15& 56& 9&144&104& 198& 2& 8 &36.4& 100& 100&  31.3&99.86&100& 30.8&91.71&100\\
16& 64& 2&109& 83& 135& 1& 2 &32.3& 100& 100&  29.3&100 &100&  29.5&96.81&100\\\bottomrule         
\end{tabular}      
\end{table*}

\begin{table*}[]
\centering
\caption{Learning from three to six seizures, and testing with the remaining seizures for patient minority.}
\label{table:Group2}
\begin{tabular}{ccclllclcclcclcc}\toprule
\multicolumn{7}{c}{\textbf{Patients information}}                                                                                           & \multicolumn{3}{c}{\textbf{LBP + HD Computing}}                                 & \multicolumn{3}{c}{\textbf{LBP + Linear SVM}}                                & \multicolumn{3}{c}{\textbf{LGP + MLP}}                                       \\\midrule
\textbf{ID} & \textbf{\begin{tabular}[c]{@{}c@{}}Elect.\\ {[\#]} \end{tabular}} & \textbf{\begin{tabular}[c]{@{}c@{}}Seiz.\\ {[\#]}\end{tabular}} &\multicolumn{3}{c}{\textbf{\begin{tabular}[c]{@{}c@{}}Seizure duration {[}s{]}\end{tabular}}}& \textbf{\begin{tabular}[c]{@{}c@{}}Trained\\
seiz. {[\#]}\end{tabular}}  & \textbf{\begin{tabular}[c]{@{}c@{}}Mean\\ delay {[}s{]}\end{tabular}} & \textbf{\begin{tabular}[c]{@{}c@{}}Spe.\\   {[}\%{]}\end{tabular}} & \textbf{\begin{tabular}[c]{@{}c@{}}Sen.\\   {[}\%{]}\end{tabular}} & \textbf{\begin{tabular}[c]{@{}c@{}}Mean\\ delay {[}s{]}\end{tabular}} & \textbf{\begin{tabular}[c]{@{}c@{}}Spe.\\   {[}\%{]}\end{tabular}} & \textbf{\begin{tabular}[c]{@{}c@{}}Sen.\\   {[}\%{]}\end{tabular}} & \textbf{\begin{tabular}[c]{@{}c@{}}Mean\\ delay {[}s{]}\end{tabular}} & \textbf{\begin{tabular}[c]{@{}c@{}}Spe.\\   {[}\%{]}\end{tabular}} & \textbf{\begin{tabular}[c]{@{}c@{}}Sen.\\   {[}\%{]}\end{tabular}} \\\cmidrule{4-6}
   &                      &                    & \textbf{Mean}                     & \textbf{Min} & \textbf{Max} &                          &                 &                    &                      &                      &                   &                    &                   &                     &                     \\ \midrule
3& 62& 14& 98& 31& 139& 3&   21.8&  94.17& 100 &15.6&87.80&100& 16.8&91.47&100 \\
7& 74& 7& 587& 154& 1002& 3& 5.0& 44& 100 &5.0&45.24&100 &      7.6&69.37&100\\
9& 92& 6& 79& 19& 100& 3&    16.2&  100& 100 &11.0&99.95&100&   21&98.61&100\\
10& 47& 13& 71& 10& 252& 3&  9.3&  99.47& 90  &9.2&99.10&90 &   13.3&99.66&70\\
12& 54& 10& 99& 80& 154& 6&  12.6&  93.10& 100 &14.5&81.61&100 &12.6&77.58&100\\
14& 49& 10& 45& 23& 93& 4&   21.7&  100& 100 &9.3&91.08&100 &   7.2&82.30&100\\\bottomrule         
\end{tabular}      
\end{table*}
We include the anonymized dataset of 16 patients of the epilepsy surgery program of the Inselspital Bern in this study for a total of 99 recordings.
Each recording consists of 3 minutes of interictal segments (immediately preceding the seizure), and the ictal segment (ranging from 10\,s to 1002\,s), followed by 3 minutes of postictal time. 
Table~\ref{table:Group1} and~\ref{table:Group2} list the number of electrodes, the number of seizures, and the seizure duration for every patient. 
To evaluate the performance of our algorithm, we use specificity, sensitivity, the number of trained seizures, and delay of detection.
The delay is measured as the total time that an algorithm takes to classify an unseen seizure after the seizure onset time point that is marked by the expert; note that it is not the implementation delay but the \emph{working} delay of an algorithm.
Based on these criteria, we observe that patients may be roughly partitioned into two groups.
For the majority of the patients (10 out of 16 in Table~\ref{table:Group1}), our algorithm quickly learns from one or two seizures, and achieves perfect (100\%) specificity and sensitivity with $k$-fold cross-validation, where $k$ is the total number of seizures minus the number of trained seizures.
Our algorithm shows 18.2\,s average delay in detection, which is well suited for several important applications considering that iEEG seizure onset often precedes clinical onset by more than 20\,s~\cite{Hirsch_Epilepsia15}.
Furthermore, a 20\,s delay from seizure onset detection may in many cases be fast enough to quench seizure activity or to prevent further spreading and secondary generalization, which is the hallmark of disabling or even life-threatening seizures.
%
%There are resource-intensive algorithms based on deep neural networks that achieve similar level of accuracy as ours, but they have been tried on small and simple dataset without capability of one- or few-shot learning~\cite{ DeepNetSeizure}.

For the remaining minority of 6 patients, listed in Table~\ref{table:Group2}, our algorithm requires more seizures (3--6) for training.
We train with the first $m$ accrued seizures listed in Table~\ref{table:Group2} and test with the remaining seizures.
For these patients we use 22 seizures for training and test with the remaining 38 seizures.
The algorithm almost maintains its top performance with 100\% sensitivity for 5 of 6 patients. 
Only one patient causes a low specificity.
It is worthwhile to discuss two types of variability that can occur in epileptic seizures.
On one hand, seizures may start focally and remain focal (i.e. restricted to one cerebral hemisphere) or they may secondary generalize and involve both cerebral hemispheres. 
Importantly for this kind of variability the iEEG patterns emerging at seizure onset are very similar and thus seizure onsets should be rapidly detected. 
On the other hand, few patients may have seizures starting in different regions of the brain with for example some seizures beginning in the left, some in the right temporal lobe (i.e. in cases of bilateral temporal lobe epilepsy). 
Then there will be different iEEG patterns at seizure onset for the different types of seizures, and both types have to be learned by the algorithm.

We also compare our algorithm with two recent methods using local pattern transformation~\cite{LGP+MLP}: 1) A method using histograms of LBPs ($2^l$ integer features per electrode) that performs best with a linear SVM classifier; 2) Akin to LBP, a local gradient pattern (LGP) is further proposed that with an MLP neural network outperforms LBP+SVM. 
In an identical setup, the postprocessing is tuned for them accordingly.
As shown across both tables, our algorithm, on average, achieves higher specificity and sensitivity than the other methods, but with $\approx$2\,s higher detection delay on average.
The low specificity of LBP+SVM and LGB+MLP clearly limits their usage for long-time recordings.
In addition, let us optimistically assume that we could aggressively quantize their weights: SVM weights with 32-bit fixed-point still require 4--10$\times$ larger memory (depending on the number of electrodes) than our algorithm; and MLP weights with 1-bit demand 5--13$\times$ larger memory. 
\section{Conclusion}
We present a simple algorithm for one-shot learning and classification of seizures.
Our algorithm exploits LBP codes and HD computing to enable completely binary operations during training and inference.
It further provides a universal and scalable interface to analyse all iEEG recordings (36--100 electrodes) with a minimal set of parameters.
Our algorithm learns from one seizure, two seizures, and three-to-six seizures, for eight, two, and six patients, respectively. 
It outperforms the LBP+SVM and LGP+MLP for detecting 65 novel seizures with higher specificity and sensitivity, and lower memory. 
%
%\section*{Acknowledgment}

{
%\small
\footnotesize
\bibliographystyle{IEEEtran}
%\bibliography{references}

}
\end{document}